\documentclass[pre, amssymb, amsmath, nofootinbib, 10pt, twocolumn]{revtex4-1}
	
\usepackage{graphicx}
\usepackage{color}
\usepackage{verbatim}
\usepackage{subfigure}
\usepackage{amssymb}
\usepackage{amsmath}

\def\f#1{Fig.~\ref{#1}}
	
\begin{document}
	
\title{Entropy production and fluctuation theorems for active matter}
\author{Dibyendu Mandal$^{1}$ and Katherine Klymko$^{2}$ and Michael R. DeWeese$^{1, 3}$}
\affiliation{
$^1$Department of Physics, University of California, Berkeley, CA 94720, USA \\
$^2$Department of Chemistry, University of California, Berkeley, CA 94720, USA \\
$^3$Redwood Center for Theoretical Neuroscience and Helen Wills Neuroscience Institute, University of California, Berkeley, CA 94720, USA}

\begin{abstract}

Active biological systems reside far from equilibrium, dissipating heat even in  their steady state, thus requiring an extension of conventional equilibrium thermodynamics and statistical mechanics. 
In this Letter, we have extended the emerging framework of stochastic thermodynamics to active matter. 
In particular, for the active Ornstein-Uhlenbeck model, we have provided consistent definitions of thermodynamic quantities such as work, energy, heat, entropy,  and entropy production at the level of single, stochastic trajectories and derived related fluctuation relations. 
We have developed a generalization of the Clausius inequality, which is valid even in the presence of the non-Hamiltonian dynamics underlying active matter systems.
We have illustrated our results with explicit numerical studies. 

\end{abstract}
	
\maketitle


Active matter systems are composed of constitutive elements that are capable of self-propulsion. 
Either through an internal mechanism or by extracting energy from their environment, these elements exhibit self-induced motion in the absence of any externally applied force. 
Examples include solutions containing single cellular organisms such as bacteria or protozoa, synthetic colloidal systems, and vibrated monolayers of granular matter~\cite{Howse_2007, Julicher_2007, Narayan_2007, Cates_2012, Palacci_2013}. 
In fact, active matter models have been used to describe flocking, schooling, and herding behavior in animal movement~\cite{Vicsek_1995, Toner_1998, Toner_2005, Vicsek_2012, Cavagna_2014}. 
Active systems are attracting growing interest due to their relevance for understanding live biological systems and their potential applications to the design of synthetic colloidal systems with controllable properties~\cite{ Baskaran_2009, Ramaswamy_2010, Romanczuk_2012, Marchetti_2013, Yeomans_2014, Menzel_2015, Bechinger_2016}.
Moreover, they exhibit novel collective properties such as phase separation in the absence of explicit attractive interactions~\cite{Tailleur_2008, Fily_2012, Buttinoni_2013, Redner_2013, Stenhammer_2013, Stenhammer_2015, Cates_2015, Redner_2016}, rectification of random fluctuations~\cite{Angelani_2009}, and spontaneous self-organization and pattern formation~\cite{Guillaume_2010, Zuiden_2016}, which make their study of interest in its own right. 


Active matter systems constitute a new class of condensed matter systems that are inherently out of equilibrium and thereby not describable by the standard, Gibbsian framework. 
While the collective behavior of active particles has been modeled by hydrodynamic equations based on conservation and symmetry principles~\cite{Marchetti_2013}, and individual active particles by various Brownian dynamics~\cite{Bechinger_2016}, a systematic framework for the \emph{nonequilibrium} thermodynamics and statistical mechanics of active matter is still in development. 
Many of the published studies so far have focused on utilizing equilibrium thermostatic notions, often based on approximating active systems by passive systems~\cite{Lau_2009, Speck_2014, Solon_2015, Bialke_2015, Maggi_2015, Trefz_2016, Speck_2016_PRE, Nikola_2016, Fily_2016}. 
In this Letter we propose an alternative approach based on stochastic thermodynamics~\cite{Jarzynski_2011, Seifert_2012}, which is an emerging framework for the description of thermodynamics and statistical mechanics of stochastic systems far from equilibrium. 
Stochastic thermodynamics has enabled us to define thermodynamic quantities such as heat, energy, work, entropy, and entropy production at the level of individual realizations of stochastic dynamics.
Moreover, one obtains exact analytical results for the fluctuations of entropy production in the form of \emph{equalities}, as opposed to the inequalities of the second law of thermodynamics. 
These equalities are more popularly known as fluctuation relations~\cite{Jarzynski_1997, Crooks_1999, Hatano_2001, Speck_2005, Seifert_2005, Esposito_2010}.


In this Letter, we choose the active Ornstein-Uhlenbeck process (AOUP), alternatively called the Gaussian colored noise model, to illustrate our approach~\cite{Fodor_2016, Flenner_2016, Paoluzzi_2016, Szamel_2016, Nardini_2016, Marconi_2017}.
Like other active matter models, this model is known to exhibit motility-induced phase separation (MIPS)~\cite{cates2015motility}.
The AOUP model is different from many other models of active matter~\cite{Ganguly_2013, Chaudhuri_2014, Chaudhuri_2016} in that there is no explicit internal drive that forces the system out of equilibrium; the active behavior of the system arises from the nonequilibrium nature of the forces from the environment. 
In particular, the damping and fluctuating forces from the environment do not satisfy the fluctuation-dissipation relation (FDR).
This fact makes it a challenge to develop stochastic thermodynamics for the AOUP as the usual approaches rely heavily on the equilibrium nature of the environment. 
We overcome this challenge by proposing an exact mathematical mapping of the AOUP, which is an overdamped Langevin model, to a passive, underdamped Langevin model with effective reservoir forces that satisfy the FDR. 
We derive our generalizations of both the first and the second laws of thermodynamics in reference to this mapped system, the latter giving rise to a modified Clausius inequality. 
Moreover, we derive both integral and detailed fluctuation relations for entropy production. 
These in turn allow us to make exact and verifiable predictions for the behavior of the original active matter system. 


This Letter is inspired in part by a recent study of the AOUP~\cite{Fodor_2016}, in which the authors studied entropy production of the AOUP based on its path integral representation. 
We provide an improvement upon their result in that the average entropy production in our treatment (Eq.~\eqref{eq:EPDetailed}) is nonzero and positive even for an AOUP in a simple harmonic potential, which is not the case in the former treatment. 
They report zero average entropy production in such cases.
We demonstrate this difference with an explicit example at the end of this Letter. 
The crucial difference between our treatment and theirs lies in the very definition of entropy production along a trajectory: In our definition (Eq.~\eqref{eq:EPnDef}) we consider the time reversal of the dynamics in addition to the reversal of trajectories, in accordance with the framework of stochastic thermodynamics~\cite{Seifert_2012, Spinney_2012}; in~\cite{Fodor_2016} the time reversal of the dynamics had not been considered
(see Eq.~(9) in \cite{Fodor_2016}).
Moreover, we have considered a time-dependent scenario in which the potential energy of the system may vary with time --- leading to a renormalized potential energy in the underdamped dynamics, different from the original one --- a case not considered in the earlier study.  


Consider a suspension of $N$ active colloidal particles with $\mathbf{x}_i$ denoting the position of the $i$-th particle. 
In the absence of the medium, the dynamics of the particles are governed by the possibly time-dependent potential $\Phi(\mathbf{X}, t)$, where $\mathbf{X} = (\mathbf{x}_1, \mathbf{x}_2, \ldots, \mathbf{x}_N)$ refers to the configuration space of the whole system. 
There are two forces from the medium: a damping force, $- \dot{\mathbf{x}_i} / \mu$ for particle $i$, and a Gaussian random force, $\mathbf{v}_i / \mu$, the latter having the following properties~\cite{Maggi_2015}: 
	\begin{equation}
	\label{eq:ColNoise}
	\langle v_{ia} \rangle = 0, \quad \langle v_{ia}(0) v_{jb}(t) \rangle = 2 \delta_{ij} \delta_{ab} \frac{D}{\tau} e^{- |t| /\tau} ,
	\end{equation} 
for all $i, j, a$ and $b$. 
Here, the angular bracket $\langle \ldots \rangle$ denotes the noise average (\emph{i.e.}, the average with respect to many realizations of the random forces $\mathbf{v}_i$); $v_{ia}$ denotes the $a$-th component of $\mathbf{v}_i$; $\delta_{xy}$ denotes the Kronecker delta function; $\tau$ is the persistence time of the noise; and $D$ is the diffusion coefficient. 
Equation~\eqref{eq:ColNoise} implies, in particular, that the random forces felt by different particles in different directions are independent of each other. 
The Langevin equation of the $i$-th particle is given by 
	\begin{equation}
	\label{eq:AOUP}
	\dot{\mathbf{x}}_i = -\mu \boldsymbol{\nabla}_i \Phi + \mathbf{v}_i.
	\end{equation}
Note that it is possible to write down a Langevin equation for the noises $\mathbf{v}_i$ themselves: $\tau \dot{\mathbf{v}}_i = -\mathbf{v}_i + \sqrt{2D} \boldsymbol{\eta}_i$, where $\boldsymbol{\eta}_i$ denotes a Gaussian random noise with the properties $\langle \eta_{ia} \rangle = 0$ and $\langle \eta_{ia}(0) \eta_{jb}(t \rangle = 2 \delta_{ij} \delta_{ab} \delta(t)$ where $\delta(t)$ denotes the Dirac delta distribution. 


Because the noise force $\mathbf{v}_i / \mu$ has exponential memory whereas the damping force $- \dot{\mathbf{x}}_i / \mu$ is memoryless, the model violates the FDR, and we have to conclude that the medium is out of equilibrium. 
As pointed out in \cite{Fodor_2016}, in the limit of vanishing persistence time ($\tau \rightarrow 0$) the model reduces to an equilibrium model satisfying the FDR with respect to temperature $T \equiv D / (k_\text{B} \mu)$ where $k_{\text B}$ is the Boltzmann constant. 
Motivated by this observation, we replace $D$ in the following by $\mu k_{\text B} T$. 
We also use the notation $\beta = 1 / k_\text{B} T$. 


As remarked in the introduction, the absence of the FDR implies that we cannot take the medium to be in equilibrium and therefore cannot utilize the framework of stochastic thermodynamics as is. 
In particular, we cannot interpret the heat given to the medium by $T$ to be the change in entropy of the medium for finite persistence time $\tau$. 
Fortunately, it is possible to overcome this challenge because of a surprising property of the fluctuations of this active matter system: the overdamped AOUP can be mapped \emph{exactly} to an underdamped Langevin process where the new, effective medium (\emph{reservoir}) is in equilibrium. 
The effective underdamped process is given by (Appendix~\ref{App:UD})
	\begin{subequations}
	\label{eq:UD}
	\begin{align}
	\dot{\mathbf{x}}_i & = \frac{\mathbf{p}_i}{m}, \\
	\dot{\mathbf{p}_i}& = -\boldsymbol{\nabla}_i \Psi - \frac{\mathbf{p}_i}{\mu m} + \sqrt{\frac{2}{\mu \beta}} \boldsymbol{\eta}_i - \mu\left(\mathbf{P} \cdot \boldsymbol{\nabla}\right) \boldsymbol{\nabla}_i \Phi,
	\end{align}
	\end{subequations}
where $\mathbf{p}_i$ is the auxiliary momentum of particle $i$; $m = \tau / \mu$ is the effective mass; $\Psi = \Phi + \mu m (\partial/ \partial t) \Phi$ is an effective potential; $\mathbf{P} = ({\mathbf p}_1, \mathbf{p}_2, \ldots, \mathbf{p}_N)$ and $\mathbf{X} = (\mathbf{x}_1, \mathbf{x}_2, \ldots, \mathbf{x}_N)$ refer to the phase space of the whole system. 
Also, we have used $\boldsymbol{\nabla}_i$ to denote the gradient with respect to $\mathbf{x}_i$ and $\boldsymbol{\nabla} = (\boldsymbol{\nabla}_1, \boldsymbol{\nabla}_2, \ldots, \boldsymbol{\nabla}_N)$ to denote the spatial gradient in the phase space of the whole system. 
The damping and the noise terms, $ - \mathbf{p}_i / \mu m$ and $\sqrt{2 / \mu \beta} \boldsymbol{\eta}_i $, respectively, satisfy the FDR with respect to temperature $T$.
In the following we interpret them to be forces from the effective, equilibrium reservoir.  
The nonequilibrium nature of this mapped dynamics arise from the momentum-dependent force $\mathbf{F}_{i, m} \equiv - \mu\left(\mathbf{P} \cdot \boldsymbol{\nabla}\right) \boldsymbol{\nabla}_i \Phi$. 
In some sense, we have decomposed the forces of the nonequilibrium medium into those of an underlying equilibrium reservoir and explicit forces. 
We now develop the results of stochastic thermodynamics around this model. 


To begin we consider the first law of thermodynamics --- namely, conservation of energy. 
The energy of the system is given by $E = (1/2) P^2 / m + \Psi$, kinetic plus potential energy. 
A trajectory $\Gamma$ over the interval $[0, t]$ is defined to be the sequence of points $\Gamma = \{\mathbf{X}_{0:t}, \mathbf{P}_{0:t} \}=\{\mathbf{X}(s), \mathbf{P}(s) | 0\leq s\leq t\}$.
Work done on the system along any $\Gamma$ is given by~\cite{Sekimoto_2010}: 
	\begin{equation}
	\label{eq:WorkExp}
	W[\Gamma] =  \int_0^t \mathrm{d}s \, \left[ \frac{\partial \Psi}{\partial s} - \mu \, \left(\mathbf{P} \cdot \boldsymbol{\nabla}\right) \boldsymbol{\nabla} \Phi \circ \frac{\mathbf{P}}{m} \right],
	\end{equation}
where the first term denotes the thermodynamic work corresponding to the conservative force, $- \boldsymbol{\nabla}_i \Psi$, and the second term denotes the mechanical work done by the nonconservative force, $\mathbf{F}_{i, m}$.\footnote{In the terminology of~\cite{Jarzynski_2007}, the first term is the so called inclusive work done on the system corresponding to the time dependence of the potential $\Psi$, and the second term the exclusive work done by the momentum-dependent force $\mathbf{F}_{i, m}$.}
Here, the circle ($\circ$) denotes Stratonovich multiplication~\cite{Gardiner_2009}, the necessity of which follows from the chain rule of derivatives~\cite{Sekimoto_1997, Sekimoto_2010}. 
The heat given to the reservoir over $\Gamma$ is the amount of work done against the reservoir forces:  
	\begin{equation}
	\label{eq:Heat}
	Q^\text{res}[\Gamma] = - \int_0^t \mathrm{d}s \, \sum_{i=1}^N \left( -\frac{\mathbf{p}_i}{\tau} + \sqrt{\frac{2}{\mu \beta}} \boldsymbol{\eta}_i\right) \circ \frac{\mathbf{p}_i(s)}{m}.
	\end{equation}
The intuition behind the formula is the following: the reservoir degrees of freedom are unstructured and random, and any energy given to a random, unstructured medium should be interpreted as heat (as opposed to work)~\cite{Callen_1985}. 
Consistency of these definitions can be seen through the following relation, the first law of thermodynamics for the current system: 
	\begin{equation}
	\label{eq:Work}
	E(t) - E(0) = W[\Gamma] - Q^\text{res}[\Gamma].
	\end{equation}
The derivation is presented in Appendix~\ref{App:Work}. 


We now consider the second law of thermodynamics. 
Let $\rho({\bf X}, {\bf P}; t)$ be the phase space probability distribution of the system at any time $t$. 
If the system is at $(\mathbf{X}(t), \mathbf{P}(t))$ at time $t$, following the developments in stochastic thermodynamics we can define the stochastic entropy~\cite{Seifert_2005} of the system at $t$ to be $s(t) = - \ln{\rho}(\mathbf{X}(t), \mathbf{P}(t);t)$. 
The change in entropy of the reservoir over the interval $[0, t]$ is the heat given to it divided by $T$ as given by the Clausius formula $\beta Q^\text{res}[\Gamma]$ (Eq.~\eqref{eq:Heat}).
Unlike the usual second law of thermodynamics for passive systems, however, the total entropy production over $[0, t]$ is not just the sum of the change in the stochastic entropy of the system, $\Delta s$, and the Clausius entropy change of the medium, $Q^\text{res} / T$. 
To see this we need to first define the time reversal of the mapped process (Appendix~\ref{App:UDTR}): 
	\begin{subequations}
	\label{eq:UDTR}
	\begin{align}
	\dot{\mathbf{x}}_i & = \frac{\mathbf{p}_i}{m}, \\
	\dot{\mathbf{p}_i}& = -\boldsymbol{\nabla}_i \Psi - \frac{\mathbf{p}_i}{\mu m} + \sqrt{\frac{2}{\mu \beta}} \boldsymbol{\eta}_i + \mu \, \left(\mathbf{P} \cdot \boldsymbol{\nabla}\right) \boldsymbol{\nabla}_i \Phi,
	\end{align}
	\end{subequations} 
obtained by keeping the reservoir terms unchanged and replacing $t$ and $\mathbf{P}$ by $- t$ and $-\mathbf{P}$, respectively, in the rest of the terms. 
We also need to consider the time reversal of the phase space trajectory $\Gamma$, given by $\Gamma^\text{r} = \{\mathbf{X}^\text{r}_{0:t}, \mathbf{P}^\text{r}_{0:t}\}$ with $\mathbf{X}^\text{r}(s) = \mathbf{X}(t-s)$ and $\mathbf{P}^\text{r}(s) = - \mathbf{P}(t-s)$. 
Next, we need to consider $P(\Gamma)$, the probability of $\Gamma$ in the mapped dynamics and $P^\text{r}(\Gamma^\text{r})$, the probability of the time-reversed trajectory $\Gamma^\text{r}$ in the time-reversed dynamics (Eq.~\eqref{eq:UDTR}). 
If there is any time dependence in $\Psi$ and $\Phi$, the time dependence has to be reversed as well in the reverse process. 
Because entropy production is a measure of time-reversal symmetry breaking, entropy production $\Sigma[\Gamma]$ along $\Gamma$  is given by 
	\begin{equation}
	\label{eq:EPnDef}
	\Sigma[\Gamma] \equiv  k_\text{B} \ln{\frac{P[\Gamma]}{P^\text{r}[\Gamma^\text{r}]}}.
	\end{equation} 
Using Eqs.~\eqref{eq:UD} and \eqref{eq:UDTR}, we can derive an explicit path integral expression (Appendix~\ref{App:EPDetailed}):
	\begin{equation} 
	\label{eq:EPDetailed}
	\frac{\Sigma[\Gamma]}{k_\text{B}} = \Delta s + \beta Q^\text{res}[\Gamma] + \frac{\mu^2 \beta}{2}  \int \left( \mathrm{d} \boldsymbol{P} \cdot \boldsymbol{\nabla} \right)^2 \Phi. 
	\end{equation}
It is easy to prove that the average of $\Sigma$ is non-negative as required by the second law of thermodynamics. 
This follows from the fact that the average of $\Sigma$ can be written as $\langle \Sigma \rangle / k_\text{B} = \sum_\Gamma P[\Gamma] \ln{[P[\Gamma] / P^\text{r}(\Gamma^\text{r})]}$, which takes the form of a relative entropy with the known property that it is never negative~\cite{Cover_2006}. 
In fact, we can derive the following expression for the average entropy production (Appendix~\ref{App:CI}):
	\begin{equation}
	\label{eq:CI}
	\langle \Sigma \rangle = k_\text{B} \Delta H + \frac{\langle Q^\text{res}\rangle}{T} + k_\text{B} \mu \int_0^t \mathrm{d}s \, \langle \nabla^2 \Phi \rangle \geq 0,
	\end{equation} 
where $H[\rho] = \langle s\rangle = - \int_\mathbf{X, P} \rho \ln{\rho}$ is the Shannon entropy of the system. 
This is the central result of our paper. 
It expresses the second law of thermodynamics as a modified Clausius inequality.
Each quantity in the inequality can be calculated even for the original dynamics as the mapped system is mathematically equivalent to the AOUP. 
The inequality therefore constitutes a prediction for the original system. 


The difference between the usual Clausius inequality and our generalization of the second law described above is embodied by the last term. 
By comparison with Eq.~\eqref{eq:EPDetailed}, it is readily apparent that the last term arises from the momentum-dependent force $\mathbf{F}_{i,  m}$, though this last term is not the average work done by $\mathbf{F}_{i,  m}$ as can be seen from Eq.~\eqref{eq:WorkExp}. 
There is, indeed, an interesting origin of this term stemming from the mapped dynamics, Eq.~\eqref{eq:UD}. 
If the reservoir terms are taken out, due to $\mathbf{F}_{i,  m}$, the dynamics is not Hamiltonian and the phase space volume is not conserved. 
The last term in Eq.~\eqref{eq:CI} is the integral of the average phase space contraction rate due to $\mathbf{F}_{i,  m}$ (Appendix~\ref{App:Contraction}). 
For deterministic thermostats, this can be interpreted as the entropy production~\cite{Andrey_1985, Cohen_1998, Dorfman_1999, Ruelle_2003}.
This is the same quantity that appeared in a recent generalization of the Jarzynski equality for non-Hamiltonian dynamics~\cite{Mandal_2016_PRE}. 
Moreover, recent developments in stochastic thermodynamics have demonstrated that the usual Clausius inequality has to be modified in the presence of feedback control~\cite{Kim_2004, Cao_2009, Sagawa_2010, Toyabe_2010, Ponmurugan_2010, Horowitz_2010, Abreu_2012, Lahiri_2012, Kundu_2012, Munakata_2012, Liu_2014, Parrondo_2015}. 
In particular, there are extra terms in the second law because the external feedback controller is not accounted for by explicit degrees of freedom. 
The momentum-dependent force $\mathbf{F}_{i,  m}$ can be seen as a spatially inhomogeneous feedback cooling operation. 
The last term in Eq.~\eqref{eq:CI} refers to this extra contribution. 
Towards the end of this Letter we will demonstrate that it is crucial to include this extra term. 
In its absence the inequality may not be satisfied. 


A major contribution from stochastic thermodynamics is the surprising result that the inequalities of the second law can be replaced by exact equalities. 
These are in a sense more refined versions of the inequalities because the latter can be derived by the application of Jensen's inequality. 
We can derive the following exact relation, the integral fluctuation relation, for entropy production in our active matter system:
	\begin{equation}
	\label{eq:IFT}
	\langle e^{- \Sigma / k_\text{B}} \rangle = \sum_\Gamma P[\Gamma] e^{- \Sigma[\Gamma] / k_\text{B}} = \sum_{\Gamma^\text{r}} P^\text{r}[\Gamma^\text{r}] = 1,
	\end{equation}
where the second relation follows from the definition of entropy production (Eq.~\eqref{eq:EPnDef}) and the third relation follows from the normalization of $P^\text{r}[\Gamma^r]$. 
Equation~\eqref{eq:IFT} is our second main result.
The modified Clausius inequality $\langle \Sigma \rangle \geq 0$ follows from the application of Jensen's inequality to Eq.~\eqref{eq:IFT}. 
In fact, there is a more detailed equality underlying Eq.~\eqref{eq:IFT} (Appendix~\ref{App:DFT}):
	\begin{equation}
	\label{eq:DFT}
	P(\Sigma = \sigma) = e^{\sigma / k_\text{B}} P^\text{r}(\Sigma^\text{r} = - \sigma),
	\end{equation}
where $P(\Sigma = \sigma)$ denotes the probability density of $\Sigma = \sigma$ in the process described by Eq.~\eqref{eq:UD} and $P^\text{r}(\Sigma^\text{r} = \sigma)$ denotes the same quantity for the reverse process (Eq.~\eqref{eq:UDTR}). 
We have to assume that the initial condition for the reverse process is obtained from the final condition of the first process by reversing the sign of the momenta. 
Equation~\eqref{eq:DFT} is the detailed fluctuation relation for entropy production.

	\begin{figure}[tbp] 
  	\includegraphics[width=0.75\columnwidth]{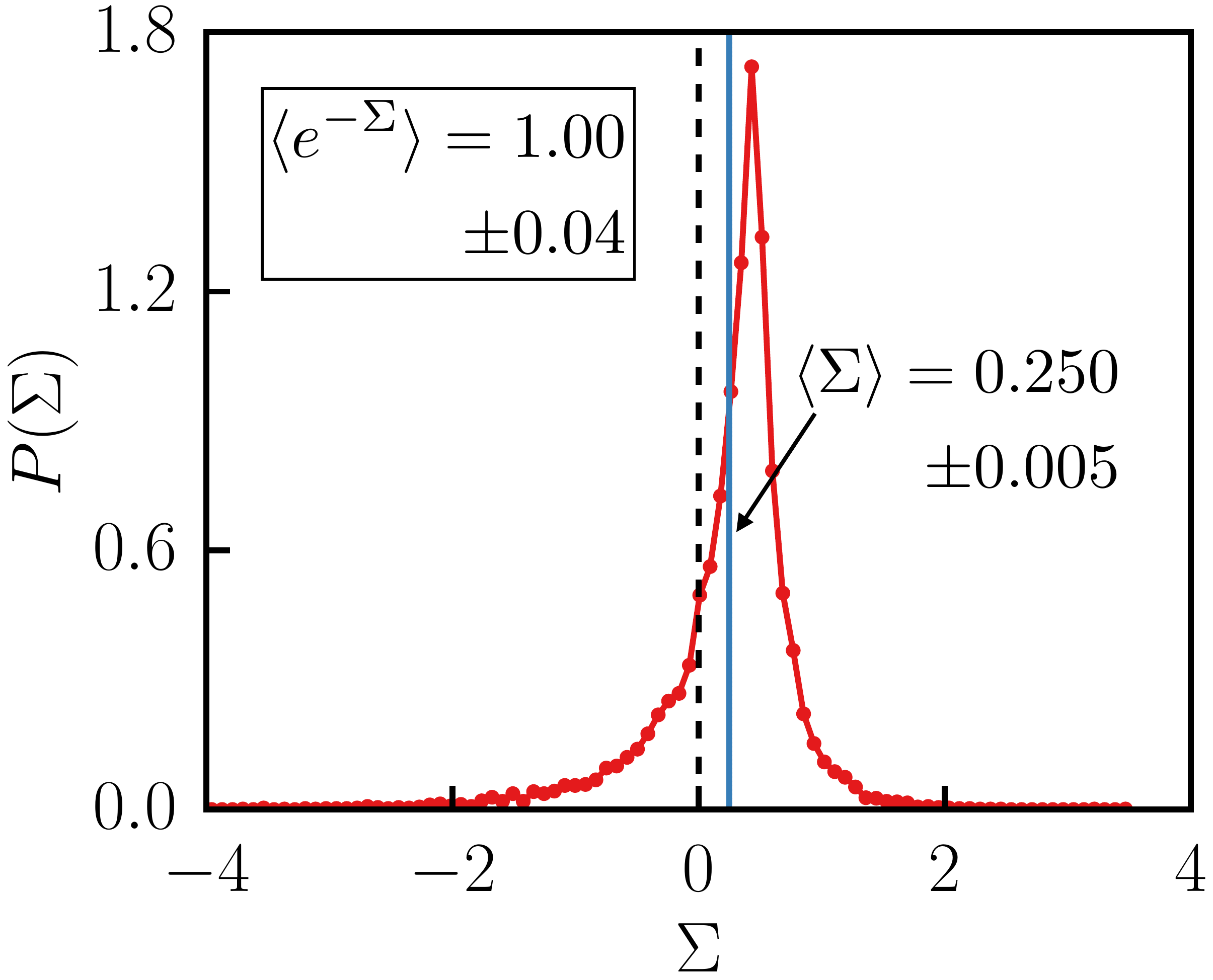}
	\caption{Numerical distribution of entropy production $\Sigma$ of a single active particle in a one-dimensional simple harmonic potential, $\Phi = x^2 / 2$. We have used temperature units to set $k_\text{B} = 1$. We see that entropy production can be negative for individual realizations. However, the average entropy production is positive, as shown by the blue, solid vertical line and the associated numerical value. Moreover, the integral fluctuation relation for entropy production (Eq.~\eqref{eq:IFT}) is satisfied.}
	\label{fig:harmonic}
	\end{figure}
 

There are three qualitatively different fluctuation relations (both integral and detailed) in stochastic thermodynamics for passive, overdamped dynamics: that of total entropy production, excess entropy production, and housekeeping heat. 
We have extended the relation for total entropy production to the AOUP.
For the sake of completeness we will address the latter two cases in the following.
The fluctuation relation for excess entropy production is a consequence of the Markovian nature of the dynamics. 
Let the steady state distribution of the dynamics in Eq.~\eqref{eq:UD} in the absence of any time dependence of $\Phi$ be $\rho^\text{s}(\mathbf{X}, \mathbf{P}; \boldsymbol{\alpha})$, where $\boldsymbol{\alpha}$ represents fixed, external parameters of the system. 
We then have the following integral fluctuation relation for transitions between any two steady states (characterized by two different values of $\boldsymbol{\alpha}$)
	\begin{equation}
	\label{eq:HSR}
	\langle e ^{ - \int_0^t \mathrm{d}s \, \dot{\boldsymbol{\alpha}} \cdot {\boldsymbol \nabla}_{\boldsymbol \alpha} \ln{\rho^s} }\rangle = 1. 
	\end{equation}
A proof of this relation follows the same basic steps as that for a related expression for passive Langevin systems, which is presented in~\cite{Hatano_2001}.  
We can rewrite the integral in the exponent as $- \Delta \ln \rho^\text{s} + \int_0^t \mathrm{d}s \, \dot{\mathbf{Z}} \circ \boldsymbol{\nabla_Z} \ln{\rho^\text{s}}$, with $\mathbf{Z}(s) = (\mathbf{X}(s), \mathbf{P}(s))$. 
The last quantity, the generalized work done against the generalized (nonequilibrium) force $- k_\text{B} T \boldsymbol{\nabla_Z} \ln{\rho^\text{s}}$, is called the excess heat $Q^\text{res, ex}[\Gamma]$. 
Equation~\eqref{eq:HSR} therefore gives the integral fluctuation relation for excess entropy production $\Sigma^\text{ex} \equiv \Delta s + Q^\text{res, ex} / T$ (in units of $k_\text{B}$). 
Unlike total entropy production, there is generally no detailed fluctuation relation for excess entropy production because the steady state distribution $\rho^\text{s}$ is generally not time-reversal symmetric, $\rho^\text{s} (\mathbf{X}, - \mathbf{P}) \neq \rho^\text{s} (\mathbf{X}, \mathbf{P})$~\cite{Lahiri_2014}.  
Nevertheless, by applying Jensen's inequality to the integral fluctuation relation~\eqref{eq:HSR} we get the following inequality satisfied by the dynamics:
	\begin{equation}
	\label{eq:CIEx}
	\Delta H + \beta \langle Q^\text{res, ex}\rangle \geq 0.
	\end{equation}
This can be a stricter bound for $\Delta H$ compared to that of Eq.~\eqref{eq:CI}.  
The concept of housekeeping heat for the current system is ambiguous due to the momentum-dependent force $\mathbf{F}_{i, m}$. 
It has been recently shown that such systems have many different notions of housekeeping heat each with its own fluctuation relation and consequent inequality~\cite{Yeo_2016}.
Given this ambiguity, we reserve a detailed discussion of the relevant results for a future study.


We now illustrate our results with a simple case study. 
Consider a single active particle in a one-dimensional simple harmonic potential, $\Phi = x^2 / 2$, where we have set the spring constant to one. 
For simplicity, we assume the other parameters, $\mu$, $\tau$, T, and the constant $k_\text{B}$, to be unity as well. 
In \f{fig:harmonic} we have plotted the probability distribution of entropy production $\Sigma$ over 0.5 units of time in the steady state of the system. 
We see that it is possible to have negative entropy production over individual trajectories, but the average over sufficiently many trajectories is always non-negative. 
In particular, the average entropy production in the current case turns out to be $\langle \Sigma \rangle = 0.250 \pm 0.005$. 
This is in contrast to the entropy production formula proposed in~\cite{Fodor_2016} where the average entropy production of active Ornstein-Uhlenbeck particles for a quadratic potential $\Phi$ is zero. 
To show the importance of the phase space contraction term in entropy production, the last term in Eq.~\eqref{eq:CI}, we have also measured the incomplete entropy production $\Sigma' = \Delta s + Q^\text{res}/ T$. 
In this case we find  $\langle \Sigma' \rangle = -0.250 \pm 0.005$.  
We see that the average is negative implying that the usual Clausius inequality is not satisfied even for an AOUP in one dimension with a simple harmonic potential.


Here we have shown how stochastic thermodynamics can be used to consistently define thermodynamic quantities for active matter~\cite{speck2016stochastic}.
We have demonstrated our method for a particular model of active matter, but we emphasize  that our approach is generalizable to other active matter systems. 
This is an important step in the construction of the thermodynamics and statistical mechanics of active matter, which will be necessary for understanding response functions and material properties of these systems.


\acknowledgments
	
The authors would like to thank Kranthi Mandadapu, Phill Geissler, Steve Whitelam, Grzegorz Szamel, Jaffar Hasnain, Jordan Horowitz, Christopher Jarzynski, Mike Hagan, Aparna Baskaran, Yaouen Fily, Fr\'ed\'eric van Wijland, Charles Frye, and Grant Rotskoff for many useful discussions. 
This work was supported in part by the U. S. Army Research Laboratory and the U. S. Army Research Office under contract W911NF-13-1-0390. K.K. acknowledges support from an NSF Graduate Research Fellowship.

	\appendix

\onecolumngrid

\section*{Supplementary materials}

\section{Derivation of Eq.~\eqref{eq:UD}}
\label{App:UD}
	
We start with the equations
	\begin{equation}
	\label{eq:AOUPApp}
	\dot{\mathbf{x}}_i = -\mu \boldsymbol{\nabla}_i \Phi + \mathbf{v}_i, \quad \tau \dot{\mathbf{v}}_i = -\mathbf{v}_i + \sqrt{2D} \boldsymbol{\eta}_i.
	\end{equation}
We define momentum by $\mathbf{p}_i = m \dot{\mathbf{x}}_i$ with $m = \tau / \mu$. 
The rate of change of $\mathbf{p}_i$ is given by 
	\begin{subequations}
	\begin{eqnarray}
	\dot{\mathbf{p}}_i & = &  m \frac{d}{dt} \dot{\mathbf{x}}_i \\
	& = & m \frac{d}{dt} \left[ -\mu \boldsymbol{\nabla}_i \Phi + \mathbf{v}_i \right] \\
	& = & m \lim_{\delta t \to 0} \frac{1}{\delta t} [ - \mu (\boldsymbol{\nabla}_i \Phi)(\mathbf{X}(t + \delta t), t + \delta t) + \mathbf{v}_i(t + \delta t) + \mu (\boldsymbol{\nabla}_i \Phi)(\mathbf{X}(t), t) - \mathbf{v}_i(t )] \\
	& = &  m \left\{ - \mu \left[ \frac{1}{m}(\mathbf{P} \cdot \boldsymbol{\nabla})+ \frac{\partial}{\partial t} \right] \boldsymbol{\nabla}_i \Phi  - \frac{\mathbf{v}_i}{\tau} + \sqrt{\frac{2 \mu}{\beta \tau^2}} \boldsymbol{\eta}_i\right\}  \\
	& = & - \mu \left( \mathbf{P} \cdot \boldsymbol{\nabla} + m \frac{\partial}{\partial t} \right) \boldsymbol{\nabla}_i \Phi - \frac{1}{\mu}\frac{\mathbf{p}_i}{m} - \boldsymbol{\nabla}_i \Phi + \sqrt{\frac{2}{\mu \beta}} \boldsymbol{\eta}_i \\
	& = & - \boldsymbol{\nabla}_i \Psi - \frac{1}{\mu}\frac{\mathbf{p}_i}{m} + \sqrt{\frac{2}{\mu \beta}} \boldsymbol{\eta}_i - \mu (\mathbf{P} \cdot \boldsymbol{\nabla}) \boldsymbol{\nabla}_i \Phi ,
	\end{eqnarray}
	\end{subequations}
with $\Psi = \Phi + m \, \mu \, \partial \Phi / \partial t$. 
In the second line we have used Eq.~\eqref{eq:AOUPApp}. 
In the fourth line we have used the chain rule of derivatives and Eq.~\eqref{eq:AOUPApp}. 
Because the noise $\mathbf{v}_i$ is colored, and not white, there is no Ito correction in the chain rule. 
In the fifth line we have used Eq.~\eqref{eq:AOUPApp} again to replace $\mathbf{v}_i$ in terms of other quantities. 
In the last line we have introduced the renormalized potential $\Psi$ to simplify the formula. 
The last line corresponds to Eq.~\eqref{eq:UD} in the main text. 

\section{Derivation of Eq.~\eqref{eq:Work}}
\label{App:Work}
	
We have
	\begin{subequations}
	\begin{align}
	\frac{d E}{dt} & = \frac{d}{dt} \left[ \frac{P^2}{2 m} + \Psi \right] \\
	& = \dot{\mathbf{P}} \circ \frac{\mathbf{P}}{m} + \frac{\partial \Psi}{\partial t} + \dot{\mathbf{X}} \circ \boldsymbol{\nabla} \Psi \\
	& = \frac{\mathbf{P}}{m} \circ \left[ - \boldsymbol{\nabla} \Psi - \frac{1}{\mu} \frac{\mathbf{P}}{m} + \sqrt{\frac{2}{\mu \beta}} \boldsymbol{\eta} - \mu (\mathbf{P} \cdot \boldsymbol{\nabla}) \boldsymbol{\nabla} \Phi \right] + \frac{\partial \Psi}{\partial t} + \frac{\mathbf{P}}{m} \circ \boldsymbol{\nabla} \Psi \\
	& =  - \frac{\delta Q^\text{res}}{dt} + \frac{\partial \Psi}{\partial t} - \mu (\mathbf{P} \cdot \boldsymbol{\nabla}) \boldsymbol{\nabla} \Phi \circ \frac{\mathbf{P}}{m} \\
	& =  - \frac{\delta Q^\text{res}}{dt} + \frac{\delta W}{dt}.
	\end{align}
	\end{subequations}
In the second line we have used the chain rule. 
Because the products are interpreted in the Stratonovich sense we do not have any Ito correction. 
In the third line we have used the equation of motion~\eqref{eq:UD} and defined $\boldsymbol{\eta} = (\boldsymbol{\eta}_1, \boldsymbol{\eta}_2, \ldots, \boldsymbol{\eta}_N)$. 
In the fourth line we have used cancellation of terms and the definition of heat given in Eq.~\eqref{eq:Heat}. 
The last line is the differential version of Eq.~\eqref{eq:Work}.

\section{Derivation of Eq.~\eqref{eq:UDTR}}
\label{App:UDTR}
	
To derive Eq.~\eqref{eq:UDTR}, we start with Eq.~\eqref{eq:UD} without the reservoir terms: 
	\begin{subequations}
	\label{eq:UDWOMed}
	\begin{eqnarray}
	\dot{\mathbf{x}}_i & = & \frac{\mathbf{p}_i}{m}, \\
	\dot{\mathbf{p}_i}& = & -\boldsymbol{\nabla}_i \Psi - \mu\left(\mathbf{P} \cdot \boldsymbol{\nabla}\right) \boldsymbol{\nabla}_i \Phi.
	\end{eqnarray}
	\end{subequations}
The rationale is that the reservoir is already in equilibrium, as it satisfies the FDR, and therefore already time-reversal symmetric in its dynamics. 
We then use the standard prescription for time reversal of replacing $t$ and $\mathbf{P}$ by $-t$ and $- \mathbf{P}$, respectively. 
We then find
	\begin{subequations}
	\label{eq:UDTRWOMed}
	\begin{eqnarray}
	\dot{\mathbf{x}}_i & = & \frac{\mathbf{p}_i}{m}, \\
	\dot{\mathbf{p}_i}& = & -\boldsymbol{\nabla}_i \Psi + \mu\left(\mathbf{P} \cdot \boldsymbol{\nabla}\right) \boldsymbol{\nabla}_i \Phi.
	\end{eqnarray}
	\end{subequations}
Finally we put the reservoir terms back in the equation to get the time-reversed dynamics in Eq.~\eqref{eq:UDTR}. 
The same approach was used by Pradhan and Seifert~\cite{Pradhan_2010} for studying the time reversal of charged Brownian dynamics in an equilibrium reservoir. 
 	
\section{Derivation of Eq.~\eqref{eq:EPDetailed}}
\label{App:EPDetailed}

We start with the definition $\Sigma[\Gamma] \equiv  k_\text{B} \ln{(P[\Gamma] / P^\text{r}[\Gamma^\text{r}])}$. 
The path probability $P[\Gamma]$ can be written as the limit
	\begin{equation}
	\label{eq:FullPropagator}
	P[\Gamma] \propto \rho(\mathbf{X}_0, \mathbf{P}_0; 0) \lim_{\delta t \rightarrow 0} \prod_{n = 0}^{t / \delta t - 1} \rho(\delta \mathbf{P}_{n} | \mathbf{X}_n, \mathbf{P}_n; t_n),
	\end{equation} 
where $(\mathbf{X}_0, \mathbf{P}_0)$ is the initial point of $\Gamma$; $\rho(\mathbf{X}_0, \mathbf{P}_0; 0)$ is the initial probability density; and $\rho(\delta \mathbf{P}_{n} | \mathbf{X}_n, \mathbf{P}_n; t_n)$ is the probability density of momentum increment $\delta \mathbf{P}_n$ over the time interval $(t_n, t_n + \delta t)$, for $t_n = n \delta t$, given the position $\mathbf{X}_n$ and momentum $\mathbf{P}_n$ at time $t_n$. 
The increment in position, $\delta \mathbf{X}_n$, is completely determined once the increment in momentum, $\delta \mathbf{P}_n$, is specified; therefore, we do not need to consider a probability density over $\delta {\mathbf X}_i$. 
Alternatively, we could have used a delta function distribution~\cite{Ganguly_2013}. 
Each probability density $\rho(\delta \mathbf{P}_{n} | \mathbf{X}_n, \mathbf{P}_n; t_n)$ can be written as the product 
	\begin{equation}
	\label{eq:IncrementProduct}
	\rho(\delta \mathbf{P}_{n} | \mathbf{X}_n, \mathbf{P}_n; t_n) = \prod_{i = 1}^N \rho(\delta \mathbf{p}_{i, n} | \mathbf{X}_{n}, \mathbf{P}_{n}; t_n)
	\end{equation}
because the noise terms of different particles are independent of each other. 
(See the discussion following Eq.~\eqref{eq:ColNoise}). 
In terms of $\rho(\delta \mathbf{p}_{i, n} | \mathbf{x}_{i, n}, \mathbf{p}_{i, n}; t_n)$, the entropy production can be written as~\cite{Spinney_2012}
	\begin{eqnarray}
	\label{eq:EPDetailedD}
	\frac{\Sigma[\Gamma]}{k_\text{B}} & = & \ln{ \left[ \frac{\rho(\mathbf{X}_0, \mathbf{P}_0; 0)}{\rho^\text{r}(\mathbf{X}^\text{r}_0, \mathbf{P}^\text{r}_0; 0)} \right]} \times \lim_{\delta t \rightarrow 0} \prod_{n = 0}^{t / \delta t - 1} \prod_{i = 1}^N \ln{ \left[ \frac{\rho(\delta \mathbf{p}_{i, n} | \mathbf{X}_{n}, \mathbf{P}_{n}; t_n)}{\rho^\text{r}(\delta \mathbf{p}_{i, n} | \mathbf{X}_{n + 1}, - \mathbf{P}_{n + 1}; t_{n+1})} \right]},\nonumber  \\
	\end{eqnarray}
with $(\mathbf{X}^\text{r}_0, \mathbf{P}^\text{r}_0) = (\mathbf{X}_t, - \mathbf{P}_t)$. 
If we choose the initial condition $\rho^\text{r}(\mathbf{X}^\text{r}_0, \mathbf{P}^\text{r}_0; 0) = \rho(\mathbf{X}^\text{r}_0, - \mathbf{P}^\text{r}_0; t)$ the first ratio on the right gives the change in stochastic entropy $\Delta s$ over $\Gamma$. 
The last term denotes the entropy production in excess of $\Delta s$.
In the following, we show that this last term is equal to $\beta Q^\text{res}[\Gamma] + \frac{\mu^2}{2}  \int_\Gamma \left[ \left( \delta \boldsymbol{P} \cdot \boldsymbol{\nabla} \right) \boldsymbol{\nabla}_i \right] \Phi$ in five steps:
	\begin{enumerate}
	\item express the Langevin equations~\eqref{eq:UD} and \eqref{eq:UDTR} in their differential forms;
	\item use the differential forms to derive explict expressions for the probability densities $\rho(\delta \mathbf{p}_{i, n} | \mathbf{x}_{i, n}, \mathbf{p}_{i, n}; t_n)$;
	\item evaluate the log ratio $\ln{\left[ \frac{\rho(\delta \mathbf{p}_{i, n} | \mathbf{x}_{i, n}, \mathbf{p}_{i, n}; t_n)}{\rho^\text{r}(\delta \mathbf{p}_{i, n} | \mathbf{x}_{i, n + 1}, - \mathbf{p}_{i, n + 1}; t_{n+1})} \right]}$ up to first order in $\delta t$;
	\item evaluate $\delta Q^\text{res}$ up to first order in $\delta t$;
	\item compare the above results to arrive at $ \prod_{i = 1}^N \ln{\left[ \frac{\rho(\delta \mathbf{p}_{i, t} | \mathbf{X}_{t}, \mathbf{P}_{t}; t)}{\rho^\text{r}(\delta \mathbf{p}_{i, t} | \mathbf{X}_{t + \delta t}, - \mathbf{P}_{t + \delta t}; t + \delta t)} \right]} = \beta \delta Q^\text{res} + \frac{\mu^2}{2} \left( \delta \boldsymbol{P} \cdot \boldsymbol{\nabla} \right)^2 \Phi$ (up to first order in $\delta t$). 
	\end{enumerate}

The differential forms of the Langevin equations~\eqref{eq:UD} and \eqref{eq:UDTR} are given by, respectively, 
	\begin{align}
	\label{eq:UDDiff}
	\delta \mathbf{x}_i = \frac{\mathbf{p}_i}{m} \delta t, & \quad \delta \mathbf{p}_i = - \boldsymbol{\alpha}_i \delta t + \sqrt{\frac{2 \delta t}{\mu \beta}} \mathbf{N}_{i, (0,1)}(t),\\
	\label{eq:UDTRDiff}
	\delta \mathbf{x}_i = \frac{\mathbf{p}_i}{m} \delta t, & \quad \delta \mathbf{p}_i = - \boldsymbol{\alpha}^\text{r}_i \delta t + \sqrt{\frac{2 \delta t}{\mu \beta}} \mathbf{N}^\text{r}_{i, (0,1)}(t),
	\end{align}
with $\boldsymbol{\alpha}_i = \boldsymbol{\alpha}_i(\mathbf{X}, \mathbf{P}, t) = \boldsymbol{\nabla}_i \Psi + \frac{\mathbf{p}_i}{\tau} + \mu (\mathbf{P} \cdot \boldsymbol{\nabla}) \boldsymbol{\nabla}_i \Phi$ and $\boldsymbol{\alpha}^\text{r}_i(\mathbf{X}, \mathbf{P}, t) = \boldsymbol{\nabla}_i \Psi + \frac{\mathbf{p}_i}{\tau} - \mu (\mathbf{P} \cdot \boldsymbol{\nabla}) \boldsymbol{\nabla}_i \Phi$. 
Here, $\mathbf{N}_{i, (0,1)}(t)$ and $\mathbf{N}^\text{r}_{i, (0,1)}(t)$ are vectors of random, normal variables. 
In writing the equations~\eqref{eq:UDDiff} and \eqref{eq:UDTRDiff} we have used the Ito interpretation.
Because the noise terms are additive, the same final results would also be obtained by using the Stratonovich interpretation.  

Using Eqs.~\eqref{eq:UDDiff} and \eqref{eq:UDTRDiff} we get the following expressions at any time $t$
\onecolumngrid
	\begin{eqnarray}
	\rho(\delta \mathbf{p}_{i} | \mathbf{X}, \mathbf{P}; t) & = & G\left[- \boldsymbol{\alpha}_i (\mathbf{X}, \mathbf{P}; t)\delta t, \frac{2 \delta t}{\mu \beta} \right]\\
	\rho^\text{r}(\delta \mathbf{p}_{i} | \mathbf{X} + \delta \mathbf{X}, - \mathbf{P} - \delta \mathbf{P}; t + \delta t) & = & G\left[- \boldsymbol{\alpha}^\text{r}_i (\mathbf{X} + \delta \mathbf{X}, - \mathbf{P} - \delta \mathbf{P}; t + \delta t) \delta t, \frac{2 \delta t}{\mu \beta} \right]
	\end{eqnarray}
where $G[a, \nu^2]$	denotes a Gaussian distribution with mean $a$ and variance $\nu^2$. 

Now we evaluate the log ratio $\ln{\left[\frac{\rho(\delta \mathbf{p}_{i} | \mathbf{X}, \mathbf{P}; t))}{\rho^\text{r}(\delta \mathbf{p}_{i} | \mathbf{X} + \delta \mathbf{X}, - \mathbf{P} - \delta \mathbf{P}; t + \delta t)}\right]}$ keeping terms up to order $\delta t$. 
 In the following, we say that two expressions $f$ and $g$ are equivalent, $ f \sim g$, if they differ by terms on the order of $(\delta t)^{1 + \epsilon}$ with $\epsilon > 0$. 
 We get, omitting some arguments for brevity, 
 	\begin{align}
	\label{eq:LogRatio}
	\ln{\left[\frac{\rho(\delta \mathbf{p}_{i} | \ldots; t))}{\rho^\text{r}(\delta \mathbf{p}_{i} | \ldots; t + \delta t)}\right]} & = \frac{\mu \beta}{4 \delta t} \left\{ \left[ \delta \mathbf{p}_i + \boldsymbol{\alpha}_i^\text{r}(\mathbf{X} + \delta \mathbf{X}, - \mathbf{P} - \delta \mathbf{P}; t + \delta t) \delta t \right]^2 - \left[ \delta \mathbf{p}_i + \boldsymbol{\alpha}_i (\mathbf{X}, \mathbf{P}, t) \delta t \right]^2 \right\} \nonumber \\
	& = \frac{\mu \beta}{4 \delta t} \left[ 2 \delta \mathbf{p}_i \delta t \cdot (\boldsymbol{\alpha}^\text{r}_i - \boldsymbol{\alpha}_i) + \left(\delta t\right)^2 \left( \boldsymbol{\alpha}^\text{r}_i \cdot \boldsymbol{\alpha}^\text{r}_i - \boldsymbol{\alpha}_i^2 \right) \right] \nonumber \\
	& = \frac{\mu \beta}{4} \left[ 2 \delta \mathbf{p}_i \cdot (\boldsymbol{\alpha}^\text{r}_i - \boldsymbol{\alpha}_i) + \delta t \left( \boldsymbol{\alpha}^\text{r}_i \cdot \boldsymbol{\alpha}^\text{r}_i - \boldsymbol{\alpha}_i^2 \right) \right].
	\end{align}
We can simplify the first term in the bracket on the right as
	\begin{align}
	& 2 \delta \mathbf{p}_i \cdot \left[\boldsymbol{\alpha}_i^\text{r}(\mathbf{X} + \delta \mathbf{X}, - \mathbf{P} - \delta \mathbf{P}; t + \delta t)  - \boldsymbol{\alpha}_i (\mathbf{X}, \mathbf{P}, t)\right] \nonumber \\
	\sim & 2 \delta \mathbf{p}_i \cdot \left[\boldsymbol{\alpha}_i^\text{r}(\mathbf{X}, - \mathbf{P} - \delta \mathbf{P}; t)  - \boldsymbol{\alpha}_i (\mathbf{X}, \mathbf{P}, t)\right] \nonumber \\
	= & 2 \delta \mathbf{p}_i \cdot \left\{ \boldsymbol{\nabla}_i \Psi - \frac{\mathbf{p}_i + \delta \mathbf{p}_i}{\tau} + \mu [(\mathbf{P + \delta P}) \cdot \boldsymbol{\nabla}] \boldsymbol{\nabla}_i \Phi - \boldsymbol{\nabla}_i \Psi - \frac{\mathbf{p}_i}{\tau} - \mu [(\mathbf{P}) \cdot \boldsymbol{\nabla}] \boldsymbol{\nabla}_i \Phi \right\} \nonumber \\
	= & 2 \delta \mathbf{p}_i \cdot \left[ - \frac{2 \mathbf{p}_i}{\tau}  - \frac{\mathbf{\delta p}_i}{\tau}  + \mu \left( \delta \mathbf{P} \cdot \boldsymbol{\nabla} \right) \boldsymbol{\nabla}_i \Phi(\mathbf{X}, t) \right].
	\end{align}
	
Similarly, we can simplify the second term as
	\begin{align}
	& \delta t \left\{ \left[\boldsymbol{\alpha}_i^\text{r}(\mathbf{X} + \delta \mathbf{X}, - \mathbf{P} - \delta \mathbf{P}; t + \delta t) \right]^2 -  \boldsymbol{\alpha}_i (\mathbf{X}, \mathbf{P}, t)^2\right\} \nonumber \\
	= & \delta t \left[\boldsymbol{\alpha}_i^\text{r}(\mathbf{X}, - \mathbf{P} - \delta \mathbf{P}; t + \delta t) -  \boldsymbol{\alpha}_i (\mathbf{X}, \mathbf{P}, t)\right] \cdot \left[\boldsymbol{\alpha}_i^\text{r}(\mathbf{X}, - \mathbf{P} - \delta \mathbf{P}; t + \delta t) +  \boldsymbol{\alpha}_i (\mathbf{X}, \mathbf{P}, t)\right] \nonumber \\
	\sim & \delta t \left(- \frac{2 \mathbf{p}_i}{\tau}\right) \cdot \left[ 2 \boldsymbol{\nabla}_i \Psi(\mathbf{X}, t) + 2 \mu \left( \mathbf{P} \cdot \boldsymbol{\nabla}\right) \boldsymbol{\nabla}_ i \Phi(\mathbf{X}, t) \right].
	\end{align}
	
Collecting results, and summing over all particles, we find
 	\begin{align}
	\label{eq:LogRatio}
	& \frac{1}{\beta} \sum_i^N \ln{\left[\frac{\rho(\delta \mathbf{p}_{i} | \ldots; t))}{\rho^\text{r}(\delta \mathbf{p}_{i} | \ldots; t + \delta t)}\right]} \nonumber \\ 
	= & \sum_i^N \frac{\mu}{4} \left\{ 2 \delta \mathbf{p}_i \cdot \left[ - \frac{2 \mathbf{p}_i}{\tau}  - \frac{\mathbf{\delta p}_i}{\tau}  + \mu \left( \delta \mathbf{P} \cdot \boldsymbol{\nabla} \right) \boldsymbol{\nabla}_i \Phi(\mathbf{X}, t) \right] + \delta t \left(- \frac{2 \mathbf{p}_i}{\tau}\right) \cdot \left[ 2 \boldsymbol{\nabla}_i \Psi(\mathbf{X}, t) + 2 \mu \left( \mathbf{P} \cdot \boldsymbol{\nabla}\right) \boldsymbol{\nabla}_ i \Phi(\mathbf{X}, t) \right]\right\} \nonumber \\
	= & \frac{\mu}{4} \left\{ 2 \delta \mathbf{P} \cdot \left[ - \frac{2 \mathbf{P}}{\tau}  - \frac{\mathbf{\delta P}}{\tau}  + \mu \left( \delta \mathbf{P} \cdot \boldsymbol{\nabla} \right) \boldsymbol{\nabla} \Phi(\mathbf{X}, t) \right] + \delta t \left(- \frac{2 \mathbf{P}}{\tau}\right) \cdot \left[ 2 \boldsymbol{\nabla} \Psi(\mathbf{X}, t) + 2 \mu \left( \mathbf{P} \cdot \boldsymbol{\nabla}\right) \boldsymbol{\nabla} \Phi(\mathbf{X}, t) \right]\right\} \nonumber \\
	= & - \delta \mathbf{P} \cdot \frac{\mathbf{P}}{\tau / \mu} - \frac{( \delta \mathbf{P})^2}{2 \tau / \mu} + \frac{\mu^2}{2} (\delta \mathbf{P} \cdot \boldsymbol{\nabla})^2 \Phi - \left[ \boldsymbol{\nabla} \Psi + \mu \left( \mathbf{P}  \cdot \boldsymbol{\nabla}\right) \boldsymbol{\nabla} \Phi \right] \cdot \frac{\mathbf{P}}{\tau/\mu} \delta t.
	\end{align}
	
An expression for $\delta Q^\text{me}$ can be obtained through
	\begin{align}
	\delta Q^\text{res} & = - \left( - \frac{\mathbf{P}}{\tau} + \sqrt{\frac{2}{\mu \beta}} \boldsymbol{\eta} \right) \circ \delta \mathbf{X} \nonumber \\
	& = - \left[ \dot{\mathbf{P}} + \boldsymbol{\nabla} \Psi + \mu \left( \mathbf{P}  \cdot \boldsymbol{\nabla}\right) \boldsymbol{\nabla} \Phi \right] \circ \delta \mathbf{X} \nonumber \\
	& \sim - \delta \mathbf{P} \circ \frac{\mathbf{P}}{\tau / \mu} - \left[ \boldsymbol{\nabla} \Psi + \mu \left( \mathbf{P}  \cdot \boldsymbol{\nabla}\right) \boldsymbol{\nabla} \Phi \right] \cdot \delta \mathbf{X} \nonumber \\
	& = - \delta \mathbf{P} \cdot \frac{\mathbf{P}}{\tau/\mu} - \frac{(\delta \mathbf{P})^2}{2 \tau / \mu} - \left[ \boldsymbol{\nabla} \Psi + \mu \left( \mathbf{P}  \cdot \boldsymbol{\nabla}\right) \boldsymbol{\nabla} \Phi \right] \cdot \frac{\mathbf{P}}{\tau/\mu} \delta t.
	\end{align}
	
Combining the last two equations, we get the desired result: $ \prod_{i = 1}^N \ln{\left[ \frac{\rho(\delta \mathbf{p}_{i, t} | \mathbf{X}_{t}, \mathbf{P}_{t}; t)}{\rho^\text{r}(\delta \mathbf{p}_{i, t} | \mathbf{X}_{t + \delta t}, - \mathbf{P}_{t + \delta t}; t + \delta t)} \right]} = \beta \delta Q^\text{res} + \frac{\mu^2 \beta}{2} \left( \delta \boldsymbol{P} \cdot \boldsymbol{\nabla} \right)^2 \Phi$.

\section{Derivation of Eq.~\eqref{eq:CI}}
\label{App:CI}

We need to show $\left\langle \frac{\mu^2 \beta}{2} \left( \delta \boldsymbol{P} \cdot \boldsymbol{\nabla} \right)^2 \Phi | \mathbf{X, P}; t \right\rangle \sim \mu \, \delta t \, \nabla^2 \Phi$.
This is best done with indicial notation:
	\begin{subequations}
	\begin{align}
	\left\langle \frac{\mu^2 \beta}{2} \left( \delta \boldsymbol{P} \cdot \boldsymbol{\nabla} \right)^2 \Phi | \mathbf{X, P}; t  \right\rangle = & \, \frac{\mu^2 \beta}{2} \sum_{i, j, a, b} \left\langle \delta p_{ia} \delta p_{jb} \right\rangle \partial_{ia} \partial_{jb} \Phi \\
	\sim & \,  \frac{\mu^2 \beta}{2} \sum_{i, j, a, b} \delta_{ij} \delta_{ab} \frac{2 \delta t}{\mu \beta} \partial_{ia} \partial_{jb} \Phi \\
	= & \, \mu \,  \delta t \,  \nabla^2 \Phi.
	\end{align}
	\end{subequations}
In the first line we have written the left hand side in indicial notation and used the fact that derivatives of $\Phi$ are known quantities if $\mathbf{X}$ is known; in the second line we have used Eq.~\eqref{eq:UDDiff} and the fact that the noise is independent in different particles and directions; and in the last line we have used the properties of Kronecker delta function to simplify the summation. 
	
\section{Phase space contraction}
\label{App:Contraction}

The phase space contraction rate is given by
	\begin{subequations}
	\begin{align}
	& \left( \boldsymbol{\nabla}, \boldsymbol{\nabla}_{\mathbf P} \right) \cdot \left\{ - \frac{\mathbf{P}}{m}, \boldsymbol{\nabla} \Psi + \left[\mu (\mathbf{P} \cdot \boldsymbol{\nabla}) \boldsymbol{\nabla} \Phi \right] \right\} \nonumber \\
	= & \boldsymbol{\nabla}_{\mathbf P} \cdot \left[\mu (\mathbf{P} \cdot \boldsymbol{\nabla}) \boldsymbol{\nabla} \Phi \right] \\
	= & \mu \sum_{ijab} \partial_{p_{ia}} (p_{jb} \partial_{jb} \partial_{ia} \Phi) \\
	= & \mu \sum_{ijab} \delta_{ij} \delta_{ab} \partial_{jb} \partial_{ia} \Phi \\
	= & \mu \nabla^2 \Phi,
	\end{align}
	\end{subequations}
where in the third and fourth lines we have used indicial notation. 

\section{DFT for entropy production}
\label{App:DFT}

Following are the steps in the proof of the detailed fluctuation relation for entropy production: 
	\begin{subequations}
	\label{eq:DFT_Der}
	\begin{align}
	P(\Sigma = \sigma) & = \sum_\Gamma P[\Gamma] \delta(\Sigma[\Gamma] - \sigma) \\
	& = \sum_{\Gamma^\text{r}} P^\text{r}[\Gamma^\text{r}] e^{\Sigma[\Gamma]/ k_\text{B}} \delta(\Sigma[\Gamma] - \sigma) \\
	& = e^{\sigma / k_\text{B}} \sum_\Gamma P^\text{r}[\Gamma] \delta(\Sigma^\text{r}[\Gamma^\text{r}] + \sigma) \\
	& = e^{\sigma / k_\text{B}} P^\text{r}(\Sigma^\text{r} = - \sigma).
	\end{align}
	\end{subequations}
The first line is just the definition of the left hand side in path integral notation; the second line follows from Eq.~\eqref{eq:EPnDef}; the third line follows from the initial condition $\rho^\text{r}(\mathbf{X}^\text{r}_0, \mathbf{P}^\text{r}_0; 0) = \rho(\mathbf{X}^\text{r}_0, - \mathbf{P}^\text{r}_0; t)$; and the last line follows from the definition of the probability density of entropy production, Eq.~(\ref{eq:DFT_Der}a), applied to the time-reversed process. 
	
\end{document}